# Cooperative Short Range Routing for Energy Savings in Multi-Interface Wireless Networks

Riccardo Fedrizzi and Tinku Rasheed
Create-Net Research Center
Trento, Italy
name.surname@create-net.org

*Abstract*— Energy efficiency in wireless networks has become an important field of research due to ever increasing energy expenditure in battery supplied mobile terminals. In this paper we present an energy efficient routing scheme for multi-standard infrastructure wireless networks based on multi-hop cooperative relaying. The aim of the proposed technique is to exploit short-range cooperation to take benefit from mobile terminals having superior links thus enable energy efficiency. Performance results show that higher data-rate yielded by cooperation can compensate the expense of higher energy due to multiple interfaces active on the same mobile terminal, making possible to observe energy efficiency gain of the system. A maximum achievable energy efficiency gain of up to 42 % was observed in our simulations when using the cooperative short range routing technique.

## I. INTRODUCTION

The unprecedented expansion of broadband communication networks have led to a significant increase in energy consumption of communication networks. Current 4G/5G vision envisages higher data rates and multi-standard radio interfaces (LTE, WiFi, DVB-H, Bluetooth, etc) to provide users with flexible connectivity. However, despite the recent research efforts, current state of art energy saving technologies cannot avoid the envisaged '*energy trap*' of 4G handset devices. Hence, there is continuous potential and interest for new strategies to address all aspects of power efficiency from the user devices to the core network infrastructure and on the means by which the devices and the network interact [1].

There are recent research efforts which propose to achieve energy savings by means of optimum cooperation between the short-range interfaces. An example of cooperative network is given in Figure 1, where devices can communicate via short-range to another node instead of keeping the active direct connection to the access network. This paradigm is beneficial for different reasons: firstly, communications in the short-range are less energy costly, secondly only those devices having good connectivity to the infrastructure network will keep active links. Note that this also optimizes the long-range channel usage increasing the overall system efficiency. Moreover, cooperation can be established by exploiting devices having both good channel conditions and good battery life-time.

Cooperative strategies include relaying techniques implemented at the physical and MAC layer, such as cooperative beamforming [3], distributed space time coding [4] or selective schemes [5][6], where single or multiple relays are selected to collaborate on information transmission. Further, cooperative routing strategies combine the route selection approach with cooperative transmissions, where the network nodes help each other and as a result, with increasing number of network devices, the overall network performance can be improved [7][8]. But this is not always true for energy savings and it strongly depends on the network conditions and the power constraints at the mobile terminals (MT).

In this paper we address the problem of energy efficiency in multi-standard infrastructure wireless networks. We present an energy efficient routing mechanism exploiting cooperation among mobile terminals through the short-range interfaces. The routing mechanism exploits the context information available from the neighboring devices using a cooperative strategy. The proposed technique is developed on top of a pre-routing layer transparent from the IP routing and the underlying technologies. This enables the routing technique to work in heterogeneous scenarios with devices embedding multiple wireless technologies. Performance results considering mobile scenarios demonstrate that energy efficiency gains of up to 42 % can be achieved using the cooperative energy saving routing protocol when compared to standard routing and relaying approaches.

The rest of the paper is structured as follows. In Section 2, we present the network model and associated assumptions. The cooperative routing approach is presented in Section 3. In Section 4, the performance analysis and evaluation results are presented and finally Section 5 concludes the paper with insights into relevant future works.

## II. NETWORK MODEL AND ASSUMPTIONS

The scenario considered in this work is composed of multi-radio devices connected to an infrastructure network as shown in Figure 1. Each device uses one interface connected to the infrastructure network (long-range) and another interface for the cooperation (short-range). Moreover, *Class B* and *Class A* devices are considered having respectively poor and good QoS long-range links.

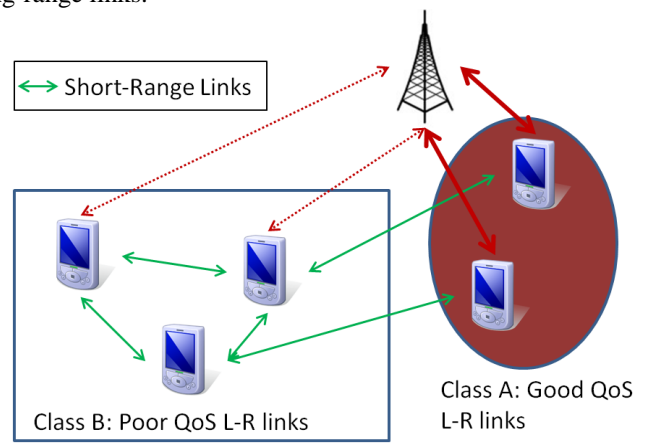

*Figure 1: Network Scenario*

The main goal of the proposed technique is to achieve energy efficiency gain performing short-range cooperation. Note that an energy efficiency gain can be expected in case of optimization of the power consumption and the data-rate. For this reason, we adopted the Energy-Per-Bit measured as J/Mb as an evaluation metric.

The main idea of the proposed method is to set up short range routes independent from the IP routing layer, in order to increase the energy efficiency through cooperation among mobile terminals (MTs) in a technology independent manner. The only interaction our model has with layers below is to retrieve the energy relevant parameters. For the above reasons we made the following assumptions: (a) MAC layers are implemented independently by our cooperation mechanisms (they only provide necessary information to our module); (b) Each MT uses the short-range interface on the same channel (how to coordinate the channels usage for the cooperation is beyond the scope of this paper).

### III. ROUTING PROTOCOL DESIGN

The proposed routing protocol is based on a modification of the distance-vector routing concept [14] in order to meet the infrastructure-based scenario. Every MT informs its neighbors about its minimum achievable energy-per-bit to reach an access gateway, through itself or through its neighbors. Considering this energy-per-bit as the route cost, the sequence of the routing decisions is taken, at every MT, considering the best neighbor to reach a gateway.

As shown in Figure 2, our proposed method is composed of two parallel states. Firstly, the Context Information Dissemination module is responsible to disseminate context information and maintain it in neighbor tables of the MTs. Secondly, the Short-Range Routing module is responsible to decide how to route the packets in order to reach the BS.

The above mentioned modules are linked with the Short-Range Energy Efficient Routing Layer which is responsible to forward to and retrieve necessary information from the above mentioned modules. The short range forwarding decisions are taken by the new layer and data packets, when received by the access point, are passed to the IP routing layer in a transparent manner. Moreover, the only interaction between the proposed technique and the underlying technologies is to retrieve necessary information to perform decisions (e.g. data-rate, power consumption levels, etc. ). For this reason, in principle, every technology can be employed as short and long-range link.

In the following, we explain the operation of the Context Information Dissemination and Short-Range Routing modules. The scope of the context information dissemination is to maintain the neighbor table for each MT participating in the cooperation. For each neighbor, the IP address for the short-range communication and the energy per bit necessary to reach the gateway (BS) through this MT are maintained regardless the number of hops needed. Every MT that is willing to cooperate sends a periodic beacon containing its IP address for short-range communications and its best $E_b$ to reach the gateway. While the IP address is related to the neighboring MT, the best energy per bit could reflect energy employed for a more than two hops route and is calculated by the procedure described below.

In a network of wireless MTs, like the scenario shown in Figure 1, let us assume that $E_b^{SR}(n,k)$ be the energy per bit to transmit via short-range interface from MT $n$ to MT $k$; $E_b^{LR}(n)$ is the energy per bit to transmit from MT $n$ to the BS employing the long-range interface; and $E_b(n)$ the lowest energy per bit found to reach the BS from MT $n$ (energy is calculated as power consumed to maintain the interface in TX state divided by the achievable data-rate).

The value $E_b(n)$ is calculated in two steps. First, we choose the best neighbor MT using Equation 1;

$$E_b(n) = \min_k\{E_b^{SR}(n,k) + E_b(k)\} \; ; \forall\, k \in NeighborTable \quad (1)$$

After that, the long-range constraint is checked as in Equation 2 in order to find which value of energy per bit to be informed via beaconing.

$$\begin{cases} E_b(n) > E_b^{LR}(n) & \rightarrow\; Tell\; E_b^{LR}(n) \\ E_b(n) < E_b^{LR}(n) & \rightarrow\; Tell\; E_b(n) \end{cases} \quad (2)$$

If a neighboring MT is found to be convenient, its related $E_b$ is communicated via beaconing, otherwise the $E_b$ related to the Long-Range interface will be exchanged.

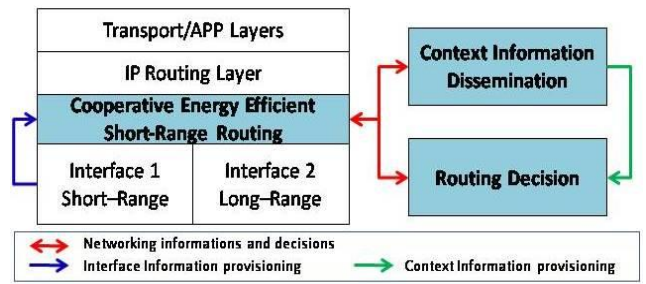

*Figure 2: cooperative short range routing approach*

In order to maintain the neighbor table, a parameter is set in order to define the beaconing period, and another parameter is set in order to define the time out for the table entries. Since broadcasted packets can be lost, every beacon updates the timestamp of the related neighbor entry while the neighbor table timeout defines how many beacons can be lost before to remove a neighbor from the list.

In this following, we explain how the context information is exploited for the routing decisions.

Suppose the MT $n$ is receiving a data packet. A two-steps procedure is run in order to decide how to route the packet. Firstly, like in the *Context Information Dissemination* stage, we find $E_b(n)$ with Equation 1. Secondly, checking the long range constraint (Equation 3), we decide whether to route the packet toward a neighbor MT or send it directly to the AP.

$$\begin{cases} E_b(n) > E_b^{LR}(n) & \rightarrow\; Send\; via\; Long\; Range \\ E_b(n) < E_b^{LR}(n) & \rightarrow\; Send\; via\; Short\; Range\; (node\; n) \end{cases} \quad (3)$$

Since the above procedure is run upon every data packet reception, the packet can be relayed several times and the route is dynamically selected. For sake of clarity, in the following we present an example showing how the proposed technique works.

#### A. Numerical Example

In Figure 3, a simple scenario is presented in which the green and red arrows represent short-range and long-range links respectively with the related costs of transmission representing the energy per bit. Moreover, in the table attached with Figure 3

the best routing choice for each MT is shown in three different points in time:

*Step 1:* At the beginning, every neighbor table is empty and every MT will use long-range interface for transmission;

*Step 2:* Each MT sent its first beacon;

*Step 3:* MT-C sent its second beacon.

As we can see in *Step 2*, MT-A and MT-C will route packets toward MT-B and MT-D respectively as they find a gain with respect to their long-range cost. In *Step 3*, MT-C notified its lowest cost ($E_b^{C \to D} + E_b^{D \to BS} = 0.3 + 3$) via beaconing reached by MTs A and B. After reaching MT-C beacon, MT-A will route its packets toward MT-C instead of MT-B, since it indirectly knows that through MT-C, it is able to exploit the resources of MT-D, bringing its total cost at *3.6* J/Mb ($E_b^{B \to C} + E_b^{C \to D} + E_b^{D \to AP}$). Moreover, MT-B is aware that it can transmit toward MT-C with a cost of *3.6* J/Mb ($E_b^{B \to C} + E_b^{C \to D} + E_b^{D \to AP} = 0.3 + 0.3 + 3$). Note that MT-D has MT-C in its neighbor table, however, since MT-C redirects traffic back to MT-D, the cost is higher than the long-range TX. This shows that the loops in the routes are naturally avoided.

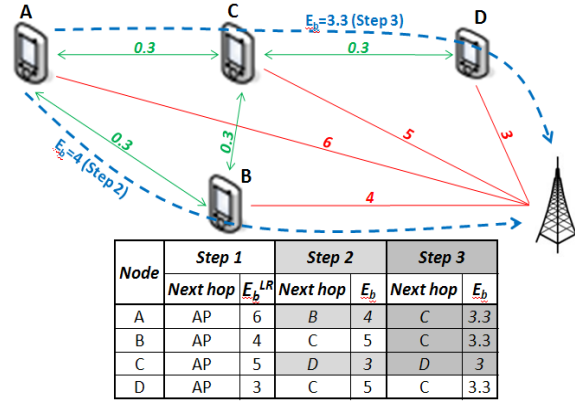

*Figure 3: example scenario describing the routing approach*

## IV. PERFORMANCE EVALUATION

In order to validate our model, intensive simulations were performed using the NS-2 network simulator. Moreover, in order to simulate multi-interface devices, we used the NS-MIRACLE library [2].

We will refer to "*Benchmark scenario*" as the reference system to understand the benefits introduced by the "*Cooperative scenario*" in which the proposed technique is employed. The peculiarities for both scenarios are described in the following.

- *Benchmark scenario:* this scenario simulates an infrastructure wireless network with homogeneous MTs equipped with only one WiMAX interface. The data flow is always sent directly to the access gateway with a constant data-rate.
- *Cooperative scenario:* this scenario simulates MTs with the proposed algorithm enabled. In order to make the cooperation possible, every device is equipped with IEEE 802.11g for the short-range cooperation and, like the benchmark scenario, WiMAX for the long-range transmission.

### A. Simulation Model

In order to simulate the proposed routing technique, we considered two simulation areas of 60 x 20 and 100 x 50 meters. For all the simulations, we set the maximum transmission range for the cooperation to 20 m, while the WiMAX BS covers the entire simulation area.

In order to have reliable results we chose to avoid having isolated MTs for which the cooperation is not possible. Also if in practice it is possible to have disjoint MTs, this could lead to difficulties in understanding system behaviors. For this reason, scenarios are created off-line for each simulation run with the following simple procedure: firstly we place MTs randomly within the simulation area and we create a graph with maximum edge length of 20 meters (maximum transmission range set for simulations), secondly we test over the created graph the any-to-any reachability running the Dijkstra algorithm [13]. If the set of positions creates a connected graph, then the scenario is admitted otherwise we try with another set of random positions.

In order to understand the benefits introduced by the cooperative energy saving routing scheme, we defined two classes of MTs as shown in Table 1. For all the MTs, data-rates of the short-range links (IEEE 802.11g) is always *54Mbps*, while for the long-range links (WiMAX) the data-rates are chosen varying the OFDM modulation. As a first instance, for the cooperative scenario we can consider *ClassB* MTs as source nodes while *ClassA* MTs as forwarding nodes.

In order to obtain the energy spent by each MT, we adopted an energy model considering:
- Time spent by each interface in TX, RX and IDLE states;
- Power consumption values for each interface in each state as shown in Table 2 [10][11].

It is important to note that the physical interfaces within the NS-MIRACLE framework do not consider the sleep state and do not have any energy efficient technique implemented. Since the *Cooperative scenario* has two active interfaces, while the *Benchmark scenario* employs only one interface, we can forecast that for the *Cooperative scenario* the absolute value of the energy consumed in the system will be always higher with respect to the *Benchmark scenario*.

TABLE 1. DATA-RATES USED FOR DIFFERENT CLASSES OF MTs

|  | "*ClassA*" MT | "*ClassB*" MT |
|---|---|---|
| **Short range Data-Rate (WiFi)** | 54 Mbps | 54 Mbps |
| **Long range Modulation (WiMAX)** | 74 Mbps | 16 Mbps |

TABLE 2. POWER CONSUMPTION VALUES FOR EACH INTERFACE STATE W

| Technology | TX | RX | IDLE |
|---|---|---|---|
| WiFi | 0.890 | 0.890 | 0.256 |
| WiMAX | 2.409 | 1.485 | 0.660 |

The aim of the simulations presented in the following is to understand if the goodput gain achieved by the cooperation can improve the energy efficiency. For this reason, the energy efficiency gain obtained in this work can be considered as a lower bound of the real obtainable gain. The overall system energy efficiency is calculated as in Equation 4.

$$E_{eff} = \frac{1}{R} \Sigma_1^R \frac{\Sigma_1^N E_C(n,r)}{\Sigma_1^N D_r(n,r)} \qquad (4)$$

where:
- *R* is the number of runs for averaging results
- *N* is number of MTs in the network
- $E_c$ is the total energy consumed J by the device *n* during simulation *r*.

- $D_r$ is the received amount of data traffic in Mbits at the Base Station sent from MT $n$ during simulation $r$.

In the following subsections, we will present results for the following cases: (i) simulations varying the number of MTs in the network, (ii) simulations varying the amount of traffic sent from the sources, and finally (iii) simulations under mobility.

For every set of parameters the simulation is repeated for *10* times, each run during 100 seconds. In order to generate traffic, we used a constant bit rate (CBR) traffic generator and all the MTs within the scenario transmit data to the BS with the same packet rate. The packet size is set to *1024 B* for all the simulations. Finally, for the context information dissemination part, we considered a beaconing period of *5* seconds for all the MTs.

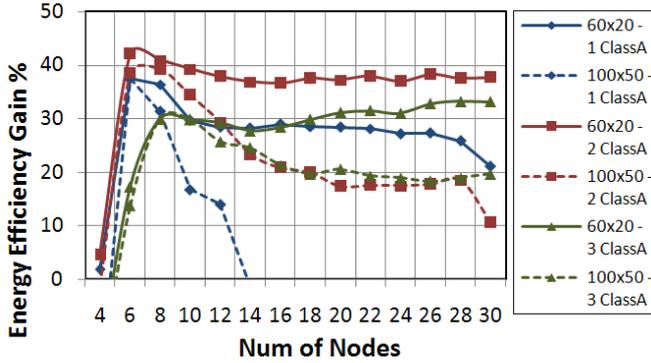

*Figure 4: energy efficiency gain varying number of Class A nodes*

### B. Simulations varying the number of MTs

In this scenario, the CBR packet frequency is set to *3000 pkts/sec* and all the MTs are transmitting to the BS. We varied the total number of MTs, keeping constant the number of *ClassA* MTs.

In Figure 4, we can see results varying the total number of MTs for different numbers of *ClassA* MTs. For very low number of nodes, the gain strongly decreases in all the cases. This is due to the fact that, the ratio between the number of *ClassA* and *ClassB* MTs decreases making it more difficult to observe any gain through cooperation. On the other hand, the observed gain, after reaching a maximum, decreases when increasing the number of *ClassB* MTs. This suggests that there is an "optimal" number of MTs that can be relayed by MTs with good connectivity to the BS.

Within the number of MTs under investigation, better gain can be achieved for 2 *ClassA* MTs in almost all the cases. As expected, when the number of relaying MTs is too low, the algorithm suffers by the fact that the L-R technology of the relaying MTs becomes the bottleneck of the system. On the other hand, a higher number of MTs with good L-R connection allows the system to successfully relay more MTs with bad L-R connection.

Observing the results for more than 20 MTs, better results are seen with 3 *ClassA* MTs in the 100x50 area, while in the smaller simulation area (60x20), the gain is however increasing towards the gain observed for 2 *ClassA* MTs. The above considerations suggest that, given a scenario, there is a particular ratio between relaying and source MTs for which the cooperation is more advisable.

Considering the results for varying topology, we can observe that with a low number of MTs in the scenario, the gain is almost similar for different simulation areas. On the other hand, in wide areas, it is easier to partition the network because of bottlenecks due to a lower connection degree in the short-range network. Therefore, with a lower number of cooperating nodes, the MT cooperation and the achievable energy gain will be heavily affected.

### C. Simulations varying the traffic

Simulations varying the amount of traffic in the network are discussed here. In Figure 5 we refer with BMK to the benchmark scenario and with CESR to our cooperative energy saving approach. In these simulations, we considered 20 MTs in the network; 4 ClassA MTs and 16 ClassB MTs. Results are shown for 60 x 20 m and 100 x 50 m areas. At first instance, we can note that the gain in the dense area (maximum 24 %) is always better that in the sparse area (maximum 17.5 %). This behavior confirms that with lower MT density, the gain is affected by the additional energy required for cooperation over multiple hops.

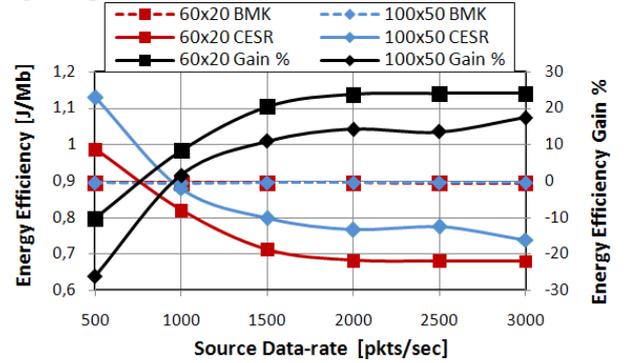

*Figure 5: Energy efficiency gain with varying source data rates (20 nodes: 4 ClassA nodes & 16 Class B nodes)*

From the results, in terms of energy efficiency for the benchmark scenario, lack of variation is observed since the WiMAX channel is congested as the network capacity limit is reached. In fact, for the benchmark scenario the average goodput is constant at *0.82 Mb/s* and the average energy consumed is also constant at *71.7 J*.

The energy saving approach exploiting cooperation with MTs having good connection to the BS is able to increase the network capacity and thus reach a gain in terms of energy efficiency thanks to a higher goodput. For high packet rate (i.e. 2000 to 3000 pkts/sec) the gain for the cooperative scheme becomes almost constant. In this case, the network reaches its capacity also with the cooperative scheme, with a constant goodput at *1.74 Mb/s*.

When the traffic is low (i.e. *500 pkts/sec*), we observe reduced gain. In this case the gain in terms of goodput (*58 %* for 60 x 20 area and 35 % for 100 x 50 area) is not able to compensate the increment of the energy consumption for the cooperation given the fact that we have to keep two active interfaces in every MT. In fact, for *500 pkts/sec* the average energy consumption in the 60 x 20 area is around *71.7 J* for the benchmark scenario, while *124.9 J* for the CESR scheme.

Note that within the ns2 framework no energy saving mechanism is implemented, thus inactive interfaces always stay in IDLE mode. This increases a lot the energy overhead for the cooperation and it is why, for low traffic (i.e. when lower goodput gains are observed), the energy efficiency gain becomes even negative.

## D. Mobility Scenario

In this section we discuss simulation results under mobility. Figure 6 shows results in terms of energy efficiency gain for the two simulation areas, having *10* transmitting MTs (2 *ClassA* MTs and 8 *ClassB* MTs). We compared the energy efficiency gain of our routing approach with the energy-aware routing algorithm (EAR), presented in [9]. For the mobility model we used the Gauss Markov mobility model [12] assuming that MTs are always inside the simulation area (i.e. MTs are bounced when they reach the area edges). In order to simulate the randomness of a walking person, α parameter of the mobility model is set equal to *0.5* for the entire simulation.

It is interesting to note that the proposed technique has comparable performances with EAR protocol when there is no mobility (figure 6). On the other hand, increasing the average speed of MTs, the cooperative energy saving approach outperforms the EAR protocol. This is more evident in the 100 x 50 m area where there is a higher probability to lose a cooperative S-R link.

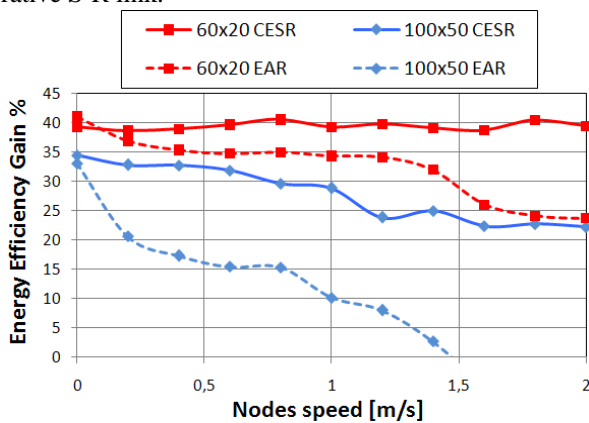

*Figure 6: energy efficiency gain with node mobility (3000 pkts/sec; 2 Class A nodes and 8 Class B nodes)*

In the 60 x 20 area, the energy efficiency gain of our approach is almost constant around *40 %*. In terms of energy consumption, the long-range energy consumption is constant at *76.9 J* while the short-range energy consumption decreases from *54.1 J* to *49.7 J*. This energy drop is due to the reduced energy used in RX state for short range, meaning that adding mobility we have more sparse MTs. However, this does not affect the goodput which is almost constant around *4 Mb/s* with slight deviations of *0.04 Mb/s*. This means that no cooperation fault occurs due to outdated neighbor entries.

On the other hand, for *100x50* simulation area, we can observe that the CESR gain is affected by mobility. In this case, for an average speed equal to *0* and *2 m/s* we have a goodput of *3.62* and *3.16 Mb/s* respectively. With more hops between a source and the BS, more time is needed in order to keep the neighbor table updated causing some non-optimal decisions by the CESR algorithm.

## V. CONCLUSIONS

In this paper we addressed the problem of energy efficiency in multi-interface infrastructure wireless networks. We presented an energy efficient routing technique exploiting cooperation among MTs through short-range interfaces. The proposed technique is developed by introducing a pre-routing layer transparent from the IP routing and the underlying wireless technology. This potentially allows the proposed technique to work with heterogeneous network interfaces. Intensive simulations were performed in order to evaluate the proposed technique which shown a maximum reachable gain around 42%. We also remark that the underlying MAC interfaces at our disposal do not consider any kind of energy saving techniques. This, of course, adds to constraints in the observed gain. The achievable gain exploiting energy saving techniques will be matter of further investigation acting on both MAC and the cooperative short range routing layer (e.g.: deactivate long-range interfaces when possible or not used). Moreover, MTs battery level and load balancing strategies will be considered in order to improve the relaying strategy.

ACKNOWLEDGMENT

This work was partly supported in part by the European (FP7-2011-8) under the Grant Agreement FP7-ICT-318632 and partly by EU-FP7 ICT-248577 C2POWER project.